\documentclass[11pt,centertags,psamsfonts]{amsart}
\usepackage{enumerate,times,amssymb,amsmath,amsfonts,epic,eepic}
\usepackage[mathcal]{euscript}

\usepackage{mathptmx}

 \numberwithin{equation}{section}

\newcommand{\loc}{{\mathop{\mathrm{loc}}}}
\newcommand{\unif}{{\mathop{\mathrm{unif}}}}

\newcommand{\ii}{\mathrm{i}}
\newcommand{\e}{\mathrm{e}}

\DeclareSymbolFont{SY}{U}{psy}{m}{n}
\DeclareMathSymbol{\emptyset}{\mathord}{SY}{'306}

\newcommand{\Z}{\mathbb{Z}}
\newcommand{\ZZ}{\mathbb{Z}}

\newcommand{\R}{\mathbb{R}}
\newcommand{\RR}{\mathbb{R}}
\newcommand{\C}{\mathbb{C}}
\newcommand{\N}{\mathbb{N}}
\newcommand{\T}{\mathbb{T}}
\newcommand{\PP}{\mathbb{P}}
\newcommand{\E}{\mathbb{E}}

\newcommand{\EE}{\mathsf{E}}
\renewcommand{\L}{L}
\newcommand{\J}{J}

\newcommand{\cC}{\mathcal{C}}

\newcommand{\cL}{\mathcal{L}}

\newcommand{\cU}{\mathcal{U}}
\newcommand{\cV}{\mathcal{V}}

\renewcommand{\Im}{{\ensuremath{\mathrm{Im}}}}
\renewcommand{\Re}{{\ensuremath{\mathrm{Re}}}}
\DeclareMathOperator*{\supp}{supp}
\newcommand{\dist}{{\ensuremath{\mathrm{dist}}}}

\newcommand{\tr}{\mathrm{tr}}

\newcommand{\fH}{\mathfrak{H}}
\newcommand{\fM}{\mathfrak{M}}

\DeclareMathOperator{\Var}{\mathrm{Var}}
\DeclareMathOperator{\wn}{\mathrm{wn}}
\DeclareMathSymbol{\emptyset}{\mathord}{SY}{'306}
\DeclareMathSymbol{\oplus}{\mathord}{SY}{'305}

\newfont{\timitfont}{pplbi7t scaled 1000}
\newfont{\timitfontsmall}{pplbi7t scaled 700}
\newcommand{\tv}{\mbox{\timitfont{v}}}
\newcommand{\tvsmall}{\mbox{\timitfontsmall{v}}}

\newcommand{\spec}{{\ensuremath{\rm spec}}}

\newcommand{\tL}{{\tilde L}}

\newcommand{\tD}{{\widetilde D}_a}

\newcommand{\Tv}{T_{\tv}}
\newcommand{\Tn}{T_n}

\newtheorem{theorem}{Theorem}[section]{\bf}{\it}
\newtheorem{proposition}[theorem]{Proposition}{\bf}{\it}
\newtheorem{corollary}[theorem]{Corollary}{\bf}{\it}
{\it}{\rm}
\newtheorem{lemma}[theorem]{Lemma}{\bf}{\it}
\newtheorem{remark}[theorem]{Remark}{\it}{\rm}
\newtheorem{definition}[theorem]{Definition}{\bf}{\it}

\newtheorem{hypothesis}{Hypothesis}{\bf}{\it}

\newtheorem{introtheorem}{Theorem}{\bf}{\it}
{\bf}{\it}
{\bf}{\it}
\newtheorem{introdefinition}{Definition}{\bf}{\it}
{\bf}{\it}

\begin{document}

\title[Lipschitz continuity of the integrated density of state]{On the Lipschitz continuity of
the integrated density of states for sign-indefinite potentials}

\author[V. Kostrykin]{Vadim Kostrykin}
\address{V. Kostrykin, Fraunhofer-Institut f\"{u}r Lasertechnik \\
Steinbachstra{\ss}e 15 \\ Aachen, D-52074, Germany}
\email{kostrykin@ilt.fraunhofer.de, kostrykin@t-online.de}
\curraddr{Institut f\"{u}r Mathematik, Technische Universit\"{a}t Clausthal,
Erzstra{\ss}e 1, D-38678 Clausthal-Zellerfeld, Germany}
\email{kostrykin@math.tu-clausthal.de}

\author[I.~Veseli\'c]{Ivan Veseli\'c}
\address{I.~Veseli\'c, Fakult\"{a}t f\"{u}r Mathematik, D-09107 TU Chemnitz, Germany}
\email{ivan.veselic@mathematik.tu-chemnitz.de\newline
{URL:\,\,www.tu-chemnitz.de/mathematik/schroedinger/} }

\begin{abstract}
The present paper is devoted to the study of spectral properties of random
Schr\"{o}dinger operators. Using a finite section method for Toeplitz matrices,
we prove a Wegner estimate for some alloy type models where the single site
potential is allowed to change sign. The results apply to the corresponding
discrete model, too. In certain disorder regimes we are able to prove the
Lipschitz continuity of the integrated density of states and/or
localization near spectral edges.
\end{abstract}



\maketitle

\section{Introduction and Main Results\label{s-intro}}

In the present work we consider random Schr\"{o}dinger operators
\begin{equation}
\label{intro:1}
H_\omega := H_0 + V_\omega,\quad H_0 :=-\Delta + V_0 \quad \text{ on }\quad \fH=L^2(\R^d),
\end{equation}
where $-\Delta$ is the negative Laplacian, $V_0$ a $\Z^d$-periodic potential,
and $V_\omega$ is given by the $\Z^d$-metrically transitive random field
\begin{equation}
\label{intro:pot}
V_\omega (x) = \sum_{j\in\Z^d}  \omega_j \, u(x-j).
\end{equation}
The bounded random variables $\omega_j, j \in \Z^d$ are assumed to be
independent and identically distributed (i.i.d.). The distribution $\mu$ of
$\omega_0$ has a density $f$ of finite total variation. It is called
\emph{single site distribution}. The probability space $\Omega=(\supp
f)^{\Z^d}$ is equipped with the product measure $\PP := \otimes_{j\in\Z^d}
\mu$. The corresponding expectation is denoted by $\E$. The function
$u\colon\R^d\rightarrow\R$ is called \emph{single site potential} and is
assumed to have compact support. We assume throughout this paper that $V_0$
and $V_\omega$ are infinitesimally bounded with respect to $\Delta$ and
that the corresponding constants can be chosen uniformly in $\omega \in
\Omega$. This is ensured, for instance, if $V_0, u \in L_{\loc,
\unif}^p(\R^d)$ with  $p=2$ for $d\le 3$ and $p>d/2$ for $d\ge 4$. Here a
function $g$ is in $ L_{\loc, \unif}^p$ if there is a constant $C$ such
that $ \int_{|x-y|<1}|g(y)|^p \mathrm{d}y \leq C$ for all $x \in \R^d$.
Without loss of generality we may assume $\min \supp f=0$ and $\max \supp
f>0$, by changing the periodic background potential $V_0$ if necessary.

The present work is devoted to the study of spectral properties of
Schr\"{o}dinger operators \eqref{intro:1} with sign-indefinite single site
potentials, i.e.~with $u$'s taking on values of both signs. The main aim of
the work is to prove the Lipschitz continuity of the \emph{integrated
density of states} (IDS) as well as a linear (with respect to the energy)
finite-volume Wegner estimate. The results presented here extend those
obtained previously by the second author in \cite{Veselic-2002a}.

For energies below the spectrum of the operator $H_0$ the H\"{o}lder continuity
of the IDS with exponent arbitrary close to $1$ was proved by Hislop and
Klopp in \cite{Hislop:Klopp} for a wide class of sign-indefinite single
site potentials. In the small disorder regime, the results of
\cite{Hislop:Klopp} also apply to internal spectral gaps of $H_0$. In the
case $d=1$, Damanik, Sims, and Stolz proved in \cite{Damanik:Sims:Stolz}
that the IDS is H\"{o}lder continuous for all energies away from a discrete set.

To describe our results let us introduce some notation: $\Lambda_l(j)$
denotes the open cube
\begin{equation*}
(-l/2, l/2)^d+j\subset\R^d
\end{equation*}
of side length $l$ centered at $j\in\Z^d$. Let $\chi_j$ be the
characteristic function of $\Lambda_1(j)$. By $H_\omega^{\Lambda}$ we denote
the restriction of the operator $H_\omega$ to the set $\Lambda$ with
periodic boundary conditions on $\partial\Lambda$. Let
$\EE_{H_\omega^\Lambda}(B)$ denote the spectral projection for the operator
$H_\omega^\Lambda$ associated with a Borel set $B\subset\R$. In particular,
if $\Lambda=\Lambda_l(0)$ we will write $H_\omega^l$ and $\EE_\omega^l(B)$
instead of $H_\omega^{\Lambda}$ and $\EE_{H_\omega^{\Lambda}}(B)$,
respectively.

The IDS $N(E)$ is defined as the limit of the distribution functions
\begin{equation*}
\begin{split}
N_\omega^l(E) &:=\; l^{-d} \; \# \{n| \, n\text{-th eigenvalue of
$H_\omega^l$ is
smaller than $E$}\}\\
&=\; l^{-d}\; \tr\,\EE_\omega^l\bigl((-\infty, E) \bigr).
\end{split}
\end{equation*}
as $l$ tends to infinity. For $\PP$-almost all $\omega \in \Omega$ the limit
exists and is independent of $\omega$.

\begin{introdefinition}\label{d-stepfunction}
Let $L^p(\R^d) \ni w \ge \kappa \chi_{0}$ with  $\kappa>0 $ and $p=2$ for
$d\le 3$ and $p > d/2$ for $d\ge 4$. Let $ \Gamma$ be a finite subset of
$\Z^d$, $\#\Gamma$ the number of its elements. A function of the form
\begin{equation}
\label{u} u(x) = \sum_{k \in \Gamma} a_k \, w(x -k)
\end{equation}
will be called a \emph{generalized step-function} and the vector $a\in
\R^{\#\Gamma}$ a \emph{convolution vector}. We set $a_k = 0$ for all $k \in
\Z^d\setminus\Gamma$ and, thus, embed $a$ in $c_0(\Z^d)$, the space of all
finite sequences with elements indexed by $j\in\Z^d$. The set $\Gamma$ will
be called the \emph{support} of $a$, $\supp\ a = \Gamma$.
\end{introdefinition}

Each convolution vector generates a multi-level Laurent (i.e.~doubly
infinite Toep\-litz) matrix $A=\{a_{j-k}\}_{j,k\in\Z^d}$ with the symbol
\begin{equation*}
s_a(\theta)=\sum_{k\in\Z^d} a_k \, \e^{\ii \langle k, \theta\rangle}, \qquad
\theta=(\theta_1,\ldots,\theta_d)\in \T^d:=(-\pi,\pi]^d.
\end{equation*}

\begin{introtheorem}[Density of states]
\label{DOStheorem} Let $u$ be a generalized step function. If $d\leq 2$ and
$s_a(\theta)\neq 0$ for all $\theta\in\T^d$, then
\begin{equation}
\label{WEresultat-1} \E \{ \tr\ \EE_\omega^l( \left[E -\epsilon,E \right])
\} \leq C\, \e^E \Var(f)\;  \epsilon \ l^d, \quad \forall \, \epsilon \geq
0,
\end{equation}
where $C$ is a constant independent of $E$, $l$, $f$, and $\epsilon$.
Moreover, the density of states $n(E):=\mathrm{d} N(E)/\mathrm{d} E$ exists
for a.e.~$E\in\R$ and is locally uniformly bounded: $n(E) \leq C \e^{E_1}
\Var(f)$ for all $E \leq E_1$.
\end{introtheorem}

Here $\Var(f)$ denotes the total variation of the function $f$.

The proof of Theorem \ref{DOStheorem} heavily relies on the finite section
method (also called projection method) for multi-level Toeplitz matrices.
We expect that Theorem \ref{DOStheorem} holds in arbitrary dimension.
However, we can prove this only under an additional assumption on the
bounded invertibility of certain auxiliary Toeplitz operators, see
Corollary \ref{cor:4.2}.

Below we will prove a weaker result (Theorem \ref{WEtheorem}), namely a
finite-volume Wegner estimate with super-linear dependence on the volume of
the cube $\Lambda_l(0)$. This estimate is useful in the context of
localization but does not allow us to say anything about the continuity of
the integrated density of states.

The symbol $s_a$ is called \emph{sectorial} if there is a $\phi \in
(-\!\pi,\pi]$ such that $\Re\;(\e^{\ii\phi} s_a(\theta)) \geq 0$ for all
$\theta\in\T^d$. A fairly simple example of a sectorial symbol is provided
by the single site potential $u(x)=\chi_0(x)-\chi_0(x-1)$ for $d=1$.
Obviously, $\Re\; s_a(\theta) = 2\sin^2(\theta/2)$ $\geq 0$ which has precisely
one zero at $\theta=0$.

\begin{introtheorem}[Wegner estimate]
\label{WEtheorem} Let $u$ be a generalized step function. Assume that the
symbol $s_a$ is sectorial
and $\Re\; s_a(\theta)$ has at most finitely many zeros.
Then there is a number $b\ge 1$ such that
\begin{equation}
\label{WEresultat}
\E \{ \tr\ \EE_\omega^l( \left[E -\epsilon,E +\epsilon\right]) \}
\leq C\, \e^E \Var(f)\;  \epsilon\; l^{bd}, \quad \forall \, \epsilon \in [0,1], \quad \forall \, l \in \mathbb{N}
\end{equation}
where $C$ is a constant independent of $E$, $l$, $f$, and $\epsilon$.
\end{introtheorem}

If the symbol $s_a$ does not vanish by Theorem \ref{DOStheorem} in
dimension one and two we can even chose $b=1$.

Slightly modifying the proof one can easily extend Theorem \ref{WEtheorem}
to the case, where $s_a(\theta)$ is independent of some of the $\theta_i$-s and,
thus, $\Re\; s_a(\theta)$ may have non-isolated zero    s. We leave the
details to the reader.

Corollary \ref{cor:4.2} below implies the following

\begin{introtheorem}\label{t-oldWE}
Let $u$ be a generalized step function. Assume there is a $k\in\Gamma$ such
that
\begin{equation}\label{domin}
|a_k| > \sum_{\substack{j\in\Gamma\\ j\neq k}} |a_j|.
\end{equation}
Then the conclusion of Theorem \ref{DOStheorem} holds for all $d\geq 1$.
\end{introtheorem}

Remark that condition \eqref{domin} implies that $s_a(\theta)\neq 0$ for all
$\theta\in\T^d$, which is part of the assumption of Theorem \ref{DOStheorem}.
Under the condition that the single site distribution $\mu$ has a
density in the Sobolev space $W^{1,1}(\R)$ this result was obtained in \cite{Veselic-2002a}.

Apart from establishing the existence of the density of states our main
application of the Wegner estimate is a proof of \emph{strong
Hilbert-Schmidt dynamical localization}. This notion means that wavepackets
with energies in an \emph{energy interval} $I\subset\R$ do not spread under
the time evolution of the operator $H_\omega$, see, e.g.,
\cite{Germinet:Klein:01a,Stollmann-2001}. More precisely,
\begin{equation}
\label{sdL} \E \left\{\sup_{\|\varphi \|_\infty \leq 1}  \left\| |X|^{q/2}
\varphi( H_\omega) \EE_{H_\omega}(I) \chi_K \right\|_{HS}^2 \right\} <
\infty
\end{equation}
holds for all $q>0$. Here $\| \cdot\|_{HS}$ denotes the Hilbert-Schmidt
norm, $K\subset\R^d$ is any compact set, and $X$ denotes the operator of
multiplication with the variable $x$. For the interpretation of \eqref{sdL}
as non-spreading of wavepackets one chooses $\varphi(y) = \e^{-\ii t y}$.
In particular, in the present context strong
Hilbert-Schmidt dynamical localization implies that the spectrum of
$H_\omega$ in $I$ is almost surely pure point with exponentially decaying
eigenfunctions (exponential spectral localization).

We prove localization for sign-indefinite single site potentials of
generalized step function form -- as long as the positive part stays
dominant -- in several energy/disorder regimes. The regimes correspond
to situations where localization has been established for fixed sign
single site potentials. We list below several situations, in which this is the case.

Set $u_+(x):= \max(u(x), 0)$ and consider the auxiliary operator
\begin{equation}\label{Hplus}
H_{\omega,+} = H_0  + V_{\omega,+}\quad\text{with}\quad V_{\omega,+}(x):= \sum_{j\in\Z^d}  \omega_j \, u_+(x-j).
\end{equation}

A number $E_0\in\spec(H_{\omega,+})$ is called a spectral edge
of the operator $H_{\omega,+}$ if there exists a
$\delta>0$ such that either $(E_0-\delta,E_0)$ or $(E_0,E_0+\delta)$
belongs to the resolvent set of the operator $H_{\omega,+}$ almost surely.
In particular, the infimum of the almost sure spectrum of the operator
$H_{\omega,+}$ is a spectral edge.

\begin{hypothesis}\label{ISEass}
Assume that $E_0$ is a spectral edge of the operator $H_{\omega,+}$.
Without loss of generality we will assume that $E_0$ is a lower spectral
edge such that $\PP\{\spec(H_{\omega,+}) \cap
(E_0-\delta,E_0)\neq\emptyset\}=0$. Assume that any of the following assumptions holds:
\begin{enumerate}
\item[(i)]
$E_0$ is the infimum of the spectrum of $H_{\omega,+}$ almost surely.
\end{enumerate}
In all subsequent cases assume that the support of the density $f$ is an
interval.
\begin{enumerate}
\item[(ii)]
for some $\tau > d/2$, some $t_0>0$ and all $t\in[0,t_0]$ the single site density $f$ satisfies
\begin{equation*}
\int_0^{t} f(x)\, \mathrm{d}x \leq C\; t^\tau
\end{equation*}
with a constant $C>0$.
\item[(iii)]
$E_0$ is a \emph{Floquet regular} spectral edge of the (periodic) operator
$H_0$, i.e., there is an $a< E_0$ such that $\spec(H_0) \cap
(a,E_0)=\emptyset$ and all Floquet eigenvalues of $H_0$ reaching $E_0$ are
locally given by Morse functions. Equivalently, the IDS of the periodic
operator $H_0$ is non-degenerate at $E_0$ (see \cite{Klopp-1999} and
\cite{Veselic-2002b}).
\end{enumerate}
\end{hypothesis}

\begin{introtheorem}[Localization]
\label{locthm} Let $u$ be a generalized step function. Assume Hypothesis
{\ref{ISEass}}. Assume, in addition, that $u\in L^\infty$ and the Wegner
estimate \eqref{WEresultat} holds with some $b>0$ for all sufficiently
large $l$. Then there exists a $\gamma > 0$ and a compact interval
$I\subset\R$ containing $E_0$ such that, if the negative part $u_-$ of the
single site potential $u$ satisfies $\|u_-\|_\infty \leq \gamma$, then
strong Hilbert-Schmidt dynamical localization holds for $H_\omega$ in the
energy interval $I$. The interval $I$ contains almost surely a spectral
edge of the operator $H_\omega$.
\end{introtheorem}

Spectral localization for single site potentials of changing sign in the energy/dis\-order regimes
(i) and (ii) was established in
\cite{Veselic-2002a}. In Section 6.2~of \cite{Hislop:Klopp} localization
results for a larger class of sign-indefinite single site potentials were
announced. However, the non-positive part of the potential still has been
assumed to be small.

In \cite{Klopp-2001a} Klopp establishes a localization result in the weak
disorder regime, i.e.~for random operators $H_0 + \alpha V_\omega$ with
$\alpha>0$ sufficiently small. The result is valid for single site
potentials $u$ with changing sign, as long as $\int u \, \mathrm{d}x \neq
0$. The weak disorder regime corresponds to the case where the support of
the distribution of coupling constants is contained in a small interval.
Correspondingly, the requirement that the density $f$ has the whole real
line as its support can be interpreted as a large disorder regime. This
case was treated by Klopp in \cite{Klopp-95a}.

\smallskip

Finally let us discuss the discrete analog of the operator family
\eqref{intro:1}. It is an \emph{Anderson model} on $\ell^2(\ZZ^d)$ with a
single site potential of finite rank:
\begin{equation}
\label{e-dAM} h_\omega
= h_0 + V_\omega, \qquad V_\omega=\sum_{j\in\Z^d} \sum_k a_k P_{k+j},
\end{equation}
where $h_0$ denotes the discrete Laplacian, $P_k$ the orthogonal projection onto the $k$-th
site of the lattice $\Z^d$,
\begin{align}\label{kor:kor}
(h_0\psi)(n)= \sum_{\substack{e\in \Z^d\\ |e|_1=1}} \psi (k+e), \qquad
(P_k\psi)(n)=\begin{cases} \psi(n), & n=k,\\ 0, & \text{otherwise}.
\end{cases}
\end{align}
The rank of the single site potential obviously equals the number of
elements in the set $\supp\; a$. Note that $V_\omega$ acts as a
multiplication operator:
\begin{equation}
\label{e-MultOp}
(V_\omega \psi)(n) =\sum_{j\in \ZZ^d} \omega_j \, a_{n-j} \,  \psi(n)
\end{equation}

We define the Laurent matrix $A$ and its symbol as above. The conclusions of
Theorems \ref{DOStheorem} and \ref{WEtheorem} remain valid for operators
$h_\omega$ as in \eqref{e-dAM}. The proofs apply verbatim. In particular,
we have

\begin{introtheorem}
\label{t-discreteDOS} If $d\leq 2$ and $s_a(\theta)$ does not vanish for
all $\theta\in\T^d$, then
\begin{equation}
\label{WEresultat-disk} \E \{ \tr\ \EE_\omega^l( \left[E -\epsilon,E
\right]) \} \leq C\,  \Var(f)\;  \epsilon \ l^d, \quad \forall \,
\epsilon \geq 0,
\end{equation}
where $C$ is a constant independent of $E$, $l$, $f$, and $\epsilon$.
Moreover, the density of states $n(E):=\mathrm{d} N(E)/\mathrm{d} E$ exists
for a.e.~$E\in\R$ and is locally uniformly bounded: $n(E) \leq C \Var(f)$
for all $E \in \R$.
\end{introtheorem}

Here $\EE_\omega^l$ is the spectral projection for the operator $h_\omega$
restricted to $\Lambda_l(0)\cap\Z^d$.

Due to equality \eqref{e-MultOp} model \eqref{e-dAM} can be understood as
the usual Anderson model with single site potential of
rank one, but with correlated random coupling constants. A similar
interpretation holds for the Schr\"{o}dinger operators \eqref{intro:1}, but in
the discrete case it is particularly clear. In fact, in the proof of
Theorems \ref{WEtheorem} and \ref{t-discreteDOS} we use this dual point of
view on the potential.

\smallskip

We give an outline of the paper. Sections \ref{s-MPSA} and \ref{s-WE}
derive abstract spectral averaging and Wegner estimates, which are applied
in Section \ref{s-symb} to prove Theorems \ref{DOStheorem} and
\ref{WEtheorem}. Section \ref{LocSec} is devoted to the proof of the
Localization Theorem \ref{locthm}. In the last section we generalize the
results of Hislop and Klopp \cite{Hislop:Klopp} on the H\"{o}lder continuity of
the IDS to single site distributions of bounded total variation. Certain
auxiliary issues are deferred to two appendices.


\subsection*{Acknowledgements}
We are indebted to A.~B\"{o}ttcher, F.~Gesztesy, K.~Makarov, and R.~Schrader
for useful and stimulating discussions. We thank the  anonymous referee for
very constructive suggestions. The work of I.V.~was supported by the DFG
through the SFB 237 ``Unordnung und gro{\ss}e Fluktuationen'' and grants no.~Ve
253/1 and Ve 253/2-1. He thanks B.~Simon for hospitality at CalTech.

\section{Multi-Parameter Spectral Averaging\label{s-MPSA}}
\numberwithin{equation}{section}

We present an extension of  the well known one-parameter spectral averaging
result of Kotani-Simon \cite{Kotani:Simon:87} and Combes-Hislop
\cite{Combes:Hislop:1994}. It is a directional averaging technique which
applies to multi-parameter families of operators.

A number of further results on the spectral averaging and its applications
to random Schr\"{o}dinger operators can be found in \cite{Buschmann:Stolz:98},
\cite{Combes:Hislop:Mourre}, \cite{Combes:Hislop:Nakamura}, \cite{Gesztesy:Makarov:Naboko},
\cite{Gesztesy:Makarov:02}, \cite[Section 3]{Kostrykin:Schrader:2000c},
\cite{Simon:95}, \cite{Simon:98}, \cite{Stolz:99}.

Let $H_0$ be a  self-adjoint operator on the separable Hilbert space $\fH$.
Let $ V\geq 0$ be an infinitesimally bounded operator with respect to $H_0$
which satisfies $0\leq \kappa B^2 \leq V$ for some $\kappa>0$ and some
bounded non-negative operator $B$. Let $\EE_{H(s)}(\cdot)$ be the spectral
family for $H(s)=H_0 + s V$, $s\in\R$.

\begin{theorem}[Spectral Averaging Theorem]
\label{Combes-Hislop} For any interval $\J\subset\R$ and for any $g\geq 0$,
$g\in L^\infty(\R)$ the inequality
\begin{equation*}
\int_\R g(s)\ B \EE_{H(s)}(\J) B\ \mathrm{d}s \leq \kappa^{-1} \|g\|_\infty
|\J|
\end{equation*}
holds in operator sense.
\end{theorem}

This theorem is proven in \cite{Combes:Hislop:1994} for functions $g$ with
compact support and bounded $V$. In \cite{Hupfer:Leschke:Mueller:Warzel} it
was observed that one can extend the estimate to $g$ with unbounded support
and in \cite{Veselic:Diss} that infinitesimal relative boundedness of $V$
is sufficient.

In Appendix \ref{appendix} we give an alternative proof of this result. It
is based on the Birman-Solomyak formula and exhibits the relation of
spectral averaging to the theory of the spectral shift function.

Here is an extension of Theorem \ref{Combes-Hislop} to multi-parameter
families.

\begin{theorem}
\label{multi:av:2} Let $f\colon \RR\to[0,\infty)$ be a function of finite
total variation with compact support such that $\|f\|_1=1$. Let
$V_1,\ldots,V_n$, $n\geq 1$ be  operators in $\fH$, which are
infinitesimally bounded with respect to $H_0$. For $s=(s_1,\ldots,s_n)\in
\R^{n}$ set
\begin{equation*}
V(s)=\sum_{i=1}^n s_i V_i, \qquad H(s)=H_0 + V(s), \qquad \text{ and }
\qquad F(s) =\prod_{i=1}^n f(s_i)
\end{equation*}
Assume that there is a nontrivial vector $t=(t_1,\ldots,t_n)\in\R^{n}$ and
a bounded, non-negative operator $B$ such that for some $\kappa>0$ the inequality
\begin{equation*}
W := \sum_{i=1}^n t_i V_i \geq \kappa B^2 \geq 0
\end{equation*}
holds in operator sense. Then the operator inequality
\begin{equation}
\label{integral3} \int\limits_{\R^{n}} \mathrm{d}s\ F(s)\ B \EE_{H(s)}(\J) B\
\leq\, {\kappa}^{-1} \| t \|_{\ell^1(\Z)}\, \Var(f)\, |\J|\,
\end{equation}
holds. Here $\EE_{H(s)}$ denotes the spectral projection for the operator
$H(s)$.
\end{theorem}

Recall that $\Var(f)$ denotes the total variation of the function $f$.

\begin{remark}\rm
\ (i) \ \
In our application $f$ will play the role of a probability density.

\ (ii) \ \ If $f$ is not of bounded total variation  we can merely  bound
the left hand side of \eqref{integral3} by a term containing
$\|f\|_\infty^n\ |\supp f|^n $. Thus, in general the bound grows
exponentially in $n$.
\end{remark}

To prove Theorem \ref{multi:av:2} we need the following well-known result
(see, e.g., \cite{Ziemer}) on the mollification of functions of bounded
total variation. We sketch its proof for the reader's convenience.

\begin{lemma}\label{molli}
Let $f\colon\R\rightarrow \R_+$ be a function of bounded total variation.
Assume, in addition, that $\displaystyle\int_\R f(x)\mathrm{d}x=1$. Then
there exists a sequence $f_k\in C_0^\infty(\R)$ such that
$\displaystyle\int_\R f_k(x)\mathrm{d}x=1$ for all $k\in\N$,
\begin{equation}
\label{limit:1} \lim_{k\rightarrow\infty} \Var(f_k) = \Var(f),
\end{equation}
and
\begin{equation}
\label{limit:2} \lim_{k\rightarrow\infty} \int_\R |f_k - f| \mathrm{d}x = 0.
\end{equation}
\end{lemma}

\renewcommand{\proofname}{Sketch of the proof}

\begin{proof}
Let $\varphi(x)$ be a non-negative function in $C_0^\infty(\R)$ with $\supp
\varphi\subset [-1,1]$ such that $\int_\R \varphi(x) \mathrm{d}x =1$. For
any $\varepsilon>0$ the function
$\varphi_\epsilon(x):=\varepsilon^{-1}\varphi(x/\epsilon)$ belongs to
$C_0^\infty(\R)$ and $\int_\R \varphi_\epsilon(x) \mathrm{d}x =1$. Now
consider the mollification of $f$,
\begin{equation*}
f(x;\varepsilon):=\int_\R \varphi_\epsilon(x-y) f(y) \mathrm{d}y.
\end{equation*}
Obviously, $f(\cdot;\varepsilon)\in C_0^\infty(\R)$ and by the Fubini
theorem $\int_\R f(x;\varepsilon) \mathrm{d}x =1$. Take a sequence
$\{\varepsilon_k\}_{k\in\N}$ converging to zero and set
$f_k(x)=f(x;\epsilon_k)$. For the proof of the relations \eqref{limit:1}
and \eqref{limit:2} we refer to Theorems 1.6.1 and 5.3.5 in \cite{Ziemer}.
\end{proof}

\renewcommand{\proofname}{Proof of Theorem \ref{multi:av:2}}

\begin{proof}
Without loss of generality we may assume $t_1> 0$ and set $k=1$. Denote
\begin{equation}
\label{vektor:t} m=(1,t_2 t_1^{-1},\ldots,t_n t_1^{-1})\in\R^{n}.
\end{equation}
Let $\eta=M s$, where $M$ is an invertible $n\times n$-matrix which acts in
the following way: $\eta_1=s_1$, $\eta_i=s_i- m_i s_1$, $i=2,\ldots,n$. We
write the integral on the l.h.s.~of \eqref{integral3} as follows
\begin{equation}
\label{integral2} \int_{\R^{n-1}} \mathrm{d} \eta^\perp \int_\R \mathrm{d}
\eta_1 \; G(\eta)\ B \EE_{H(M^{-1}(\eta))}(\J) B,
\end{equation}
where ${\eta}=(\eta_1,\eta^\perp) = (\eta_1,\eta_2,\ldots,\eta_n)$ and
\begin{equation*}
G(\eta) = F(M^{-1}\eta)= f(\eta_1) \prod_{j=2}^n f(\eta_j + m_j \eta_1).
\end{equation*}
The operator $V(s)$ in the $\eta$-variables is given by
\begin{equation*}
V(s)= V(M^{-1}\eta) = \sum_{j=2}^n \eta_j V_j + \eta_1 \sum_{j=1}^n m_j V_j
= \sum_{j=2}^n \eta_j V_j + \eta_1 t_1^{-1} W.
\end{equation*}
By the assumptions on $W$, the Spectral Averaging Theorem
\ref{Combes-Hislop} applies to the integral \eqref{integral2} and shows
that it is bounded in operator sense by
\begin{equation}
\label{integral4} \frac{t_1}{\kappa} \, |\J| \;  \int_{\R^{n-1}} \mathrm{d}
\eta^\perp\ \sup_{\eta_1\in\R} G(\eta).
\end{equation}
Assume first $f\in C^1_0(\R)$. By the fundamental theorem of calculus
\begin{equation*}
\sup_{\eta_1\in\R} G(\eta) \leq \int_\R \left|\, \big(\partial_1
G\big)(\eta_1, \eta^\perp)\right| \mathrm{d} \eta_1,
\end{equation*}
where $\partial_1$ denotes the derivative with respect to the first
variable. A calculation shows
\begin{equation*}
\big(\partial_1 G\big)(\eta)
=\sum_{j=1}^n m_j f'\big((M^{-1}\eta)_j\big)
\prod_{\substack{k=1\\ k\neq j}}^n f\big((M^{-1}\eta)_k\big)
\end{equation*}
and, thus,
\begin{equation*}
\big(\partial_1 G\big)(M s) =\sum_{j=1}^n m_j f'(s_j) \prod_{\substack{k=1\\
k\neq j}}^n f(s_k).
\end{equation*}
Therefore, the integral \eqref{integral4} is bounded by
\begin{multline*}
\frac{t_1}{\kappa} \,|\J| \
\int_{\R^{n-1}}\mathrm{d} \eta^\perp \int_\R \mathrm{d}\eta_1 \,
|\partial_1 G(\eta)|
 =
\frac{t_1}{\kappa} \,|\J| \,   \, \int_{\R^n} \mathrm{d}s |\partial_1 G(M s)|
\leq
\frac{t_1}{\kappa} \, \|f'\|_1 \, \|m\|_{\ell^1(\Z)} \,|\J|,
\end{multline*}
which yields the estimate
\begin{equation*}
\int\limits_{\R^n} \mathrm{d}s\ F(s)\ B \EE_{H(s)}(\J) B\
= \int_{\R^n} \mathrm{d} \eta\, G(\eta)\, B \EE_{H(M^{-1}({\eta}))}(\J) B
\leq \frac{\|t\|_{\ell^1(\Z)} \, \|f'\|_1}{\kappa} \, |\J|.
\end{equation*}
Recall that
\begin{equation*}
\int_\R \mathrm{d} s |f'(s)| = \Var (f).
\end{equation*}
Thus, \eqref{integral3} is proven for $f\in C_0^1(\R)$.

Now let $f$ be a function of bounded total variation. By Lemma \ref{molli}
there is a sequence of $C_0^\infty$-functions $\{f_k\}$ such that
\eqref{limit:1} and \eqref{limit:2} hold. We have
\begin{equation}
\label{integral3:dif}
\begin{split}
 \int\limits_{\R^n} \mathrm{d}s\; F(s)\ B \EE_{H(s)}(\J) B\; &=
\int\limits_{\R^n} \mathrm{d}s\; \prod_{i=1}^n f_k(s_i)\ B
\EE_{H(s)}(\J) B\\ & + \int\limits_{\R^n} \mathrm{d}s\; \left[
\prod_{i=1}^n f(s_i)-\prod_{i=1}^n f_k(s_i)\right]\ B
\EE_{H(s)}(\J) B.
\end{split}
\end{equation}
A telescoping argument shows that the norm of the second integral is bounded by
\begin{equation*}
n\|B\|^2 \int_\R |f(s)-f_k(s)| \mathrm{d}s,
\end{equation*}
which by \eqref{limit:2} tends to zero as $k\rightarrow\infty$. By our
previous argument the fist integral in \eqref{integral3:dif} is bounded by
\begin{equation*}
\kappa^{-1}\| t \|_{\ell^1(\Z)}\, \Var(f_k)\  |\J|.
\end{equation*}
Applying \eqref{limit:1} completes the proof of the theorem.
\end{proof}

\section{Wegner Estimate\label{s-WE}}

In this section we prove a Wegner estimate which applies to alloy
type models as described in Section \ref{s-intro} under the additional
Hypothesis \ref{hyp4} below.

We fix some notation: For an open set $\L\subset \R^d$, $\tL$ is the set of
lattice sites $j \in \ZZ^d$ such that the characteristic function $\chi_j$
of the cube $\Lambda_1(j)$ does not vanish identically on $\L$. Set
\begin{equation*}
U(\L) = \{j\in\Z^d|\; u(x-j)\;\text{does not vanish identically on}\;
\L\}.
\end{equation*}

\begin{hypothesis}\label{hyp4}
\; {\rm (i)} \ Assume that there is a sequence $\L_n$, $n\in\N$ of open
subsets of $\RR^d$, a sequence of finite sets $\Sigma_n\subset \Z^d$,
$n\in\N$ and a number $n_0\in \N$ such that for arbitrary $n\ge n_0$ and
every $j\in\tL_n$ there is a vector $t(j,n)\in\R^{\Sigma_n}$ such that
\begin{equation*}
\sum_{k\in\Sigma_n} t_k(j,n)\ u(x-k)\ \geq\ \chi_j(x)\quad\text{for
all}\quad x\in\L_n.
\end{equation*}

{\rm (ii)} \; \; Assume $\displaystyle \sup_{n\geq n_0}
\max_{j\in\widetilde{\L}_n} \|t(j,n)\|_{\ell^1(\Sigma_n)} < \infty$.
\end{hypothesis}

In our applications we choose the $\L_n$ to be cubes or more general
polytopes and the $\Sigma_n$ to be subsets of $\Z^d$ containing $\tL_n$
(see proof of Theorem \ref{leq:2}). With this choice the boundaries
$\partial \L_n$ are sufficiently regular so that Neumann boundary
conditions are well defined.

Throughout the present and the next sections we adopt the following
convention. The symbol $H_\omega^\L$ for an arbitrary open set $\L\subset
\RR^d$ denotes the restriction of $H_\omega$ to $\L$ with Dirichlet
boundary conditions. In the special case, when $\L$ is a cube,
$H_\omega^\L$ will denote the restriction of $H_\omega$ to $\L$ either with
Dirichlet or with periodic boundary conditions. All results stated below in
Sections \ref{s-WE} and \ref{s-symb} remain valid for both types of
boundary conditions.

\begin{theorem}\label{Proposition:3.1}
Assume part \textrm{(i)} of Hypothesis {\ref{hyp4}}. Let $\J=[E_1,E_2]$
be an arbitrary interval. Then for any $n\geq n_0$
\begin{equation*}
\E\bigl\{\tr\ \EE_{H_\omega^{\L_n}}(\J) \bigr\} \leq C\ \e^{E_2} \Var(f)
\ \max_{j\in\widetilde{\L_n}} \|t(j,n)\|_{\ell^1(\Sigma_n)} \ |\J| \ \#
\tL_n
\end{equation*}
with $\# \tL_n$ the number of elements of $\tL_n$, $C$ a constant
independent of $\L_n$ and $\J$.
\end{theorem}

\renewcommand{\proofname}{Proof}

\begin{proof}
Set $\Sigma:= \Sigma_n$ and $\L:=\L_n$.
As in \cite[Section 4]{Combes:Hislop:1994} we estimate
\begin{equation*}
\begin{split}
\E\bigl\{\tr\; \EE_{H_\omega^{\L}}(\J) \bigr\} & \leq \e^{E_2}
\E\bigl\{\tr\; \EE_{H_\omega^{\L}}(\J) \e^{-H_\omega^{\L}}
\bigr\}\;
\\ & \leq\ \e^{E_2} \sum_{j\in\tL} \| \E\bigl\{\widetilde{\chi}_j\;
\EE_{H_\omega^{\L}}(\J)\; \widetilde{\chi}_j \bigr\}\|\;
\sup_{\omega\in\Omega}\; \tr\Bigl(\e^{-H_\omega^{j}}\Bigr),
\end{split}
\end{equation*}
where $\widetilde{\chi}_j={\chi}_j\;\chi_\L$ with $\chi_\L$ the
characteristic function of the set $\L$. The operator $H_\omega^{j}$ is the
restriction of $H_\omega$ onto $\Lambda_1(j)\cap\L$ with Neumann boundary
conditions. Noting that
\begin{equation*}
C\ :=\  \sup_{\omega\in\Omega}\ \tr\Bigl(\e^{-H_\omega^{j}}\Bigr)
\end{equation*}
is bounded uniformly in $j$, we obtain the inequality
\begin{equation}
\label{ref:ref} \E\bigl\{\tr\ \EE_{H_\omega^{\L}}(\J) \bigr\}\ \leq\ C\
\e^{E_2} \sum_{j\in\tL} \| \E\bigl\{\widetilde{\chi}_j\;
\EE_{H_\omega^{\L}}(\J)\; \widetilde{\chi}_j \bigr\}\|.
\end{equation}

Recall that
\begin{equation*}
H_\omega^\L
=-\Delta^\L + \chi_\L \sum_{k\in U(\L)} \omega_k u(\cdot-k)
\end{equation*}
in the sense of quadratic forms, where $\Delta^\L$ is the Laplace operator
with Dirichlet boundary conditions on $\partial\L$, if $\L$ is arbitrary,
and either periodic or Dirichlet boundary conditions, if $\L$ is a cube.

Fix all $\omega_j$ with $j\in U(\L)\setminus\Sigma$. By part (i) of
Hypothesis {\ref{hyp4}} we can apply Theorem \ref{multi:av:2} to the
multi-parameter operator family
\begin{equation*}
\{\omega_k\}_{k\in\Sigma} \mapsto \Big(-\Delta^\L +
\chi_\L\sum_{j\in U(\L)\setminus\Sigma} \omega_j u(\cdot-j) \Big)
+ \sum_{k\in \Sigma} \omega_k u(\cdot-k),
\end{equation*}
thus, obtaining for all $j \in \tL$
\begin{equation*}
\| \E\bigl\{\widetilde{\chi}_j\ \EE_{H_\omega^{\L}}(\J)\
\widetilde{\chi}_j \bigr\}\|\ \leq\ \Var(f)  \, \|t(j,n)\|_{\ell^1(\Sigma)}
\,  |\J|.
\end{equation*}
{}From this and \eqref{ref:ref} the claim follows.
\end{proof}

\section{Generalized Step Functions\label{s-symb}}

In this section we consider a class of sign-indefinite single site
potentials for which it is particularly simple to verify Hypothesis
{\ref{hyp4}}. Throughout this section we assume that the single site
potential is a generalized step functions (see
Definition \ref{d-stepfunction}).

Each convolution vector generates a multi-level Laurent (i.e.~doubly-infinite Toep\-litz)
matrix, $A=\{a_{j-k}\}_{j,k\in\Z^d}$ whose symbol will be denoted by
$s_a$,
\begin{equation*}
s_a(\theta)=\sum_{k\in\Z^d} a_k \, \e^{\ii \langle k, \theta\rangle}, \qquad
\theta=(\theta_1,\ldots,\theta_d)\in \T^d:=(-\pi,\pi]^d.
\end{equation*}
For every $i=1,\ldots,d$ we define the $i$-th winding number
\begin{equation*}
\wn_i(s_a) = \frac{1}{2\pi \ii}\int_\T \frac{\mathrm{d}}{\mathrm{d}t} \log
s_a(\theta_1,\ldots,\theta_i=t,\ldots,\theta_d)\ \mathrm{d}t.
\end{equation*}
This number is an integer independent of $\theta$. The vector
\begin{equation*}
\wn(s_a):=(\wn_1(s_a),\ldots,\wn_d(s_a))\in\Z^d
\end{equation*}
is called also the topological index of the symbol $s_a$.

A translation of the convolution vector $a$ by an arbitrary $j_0\in\Z^d$
leaves the operator $H_\omega$ unchanged up to unitary equivalence. Indeed,
\begin{eqnarray*}
V_\omega(x - j_0) = \sum_{j \in \Z^d} \omega_j \sum_{k \in \Gamma+j_0}
a'_k \ w(x - j - k),
\end{eqnarray*}
where $a'_k\ :=\ a_{k-j_0}$. Obviously, $\supp\ a' = \supp\ a + j_0$ and the
symbol of $a'$ is given by
\begin{equation*}
s_{a'}(\theta) = \e^{-\ii\langle j_0,\theta\rangle} s_a(\theta)
\end{equation*}
such that by the product rule for winding numbers
\begin{equation}
\label{umdreh}
\wn(s_{a'}) = \wn(s_a) - j_0.
\end{equation}
\smallskip

Me make now a specific choice of the sequences $\L_n$ and $\Sigma_n$.

\begin{hypothesis}
\label{hyp5} Let $\L$ be the interior of a bounded and connected polytope in
$\R^d$ whose vertices belong to $\Z^d$. Let $\cV(\L)$ denote the set of all
vertices of the closure $\overline{\L}$. Let $K_{\tv}$ denote the cone in
$\R^d$ which at $\tv\in\cV(\L)$ locally coincides with the polytope
$\overline{\L}$, i.e., there is a neighborhood $\cU\subset\R^d$ of $\tv$
such that $K_{\tv}\cap\cU = \overline{\L} \cap \cU$. Set $\Sigma_n=\{nx| \,
x \in \overline{\L}\cap \ZZ^d\}$ and choose a sequence $\L_n\subset \R^d,
n\in \N$ such that $\tL_n \subset \Sigma_n$. In particular, the last
condition is satisfied, if $\L_n$ is the union of the unit cubes centered
at sites in $\Sigma_n$. Let $\Tv$ (respectively $\Tn$) denote the
(multi-level) Toeplitz operator which is the compression of $A$ to the
subspace $\ell^1(K_{\tv} \cap \Z^d)$ (respectively $\ell^1(\Sigma_n )$).
\end{hypothesis}

\begin{theorem}
\label{thm:Kozak} Assume Hypothesis {\ref{hyp5}}. If for every
$\tv\in\cV(\L)$ the operator $\Tv$ is continuously invertible in
$\ell^1(K_{\tv} \cap \Z^d)$, then Hypothesis {\ref{hyp4}} is satisfied.
\end{theorem}

The proof of the theorem uses some results on the finite section method for
Toeplitz operators. For an accessible introduction to this subject see,
e.g., \cite{GohbergF-1974}, \cite{BoettcherS-1999} and for an detailed
account \cite{Boettcher:Silbermann}.

\begin{proof}
{}From Kozak's Theorem \cite{Kozak:73} (for $d=2$ this is Theorem 8.57 in
\cite{Boettcher:Silbermann}) it follows that for sufficiently large $n\geq
n_0$ the operators $\Tn$ on $\ell^1(\Sigma_n)$ are continuously invertible
and the norm $\|\Tn^{-1}\|_{1,1}$ is bounded uniformly in $n$. Thus, for
every $n\geq n_0$ and any $j\in\Sigma_n$ the equation
\begin{equation}
\label{e-Loesung}
\Tn\ t(j,n) = \delta_j
\end{equation}
has a solution.
Here $\delta_j$ denotes the vector whose $j$-th component equals one and all others vanish.
Thus,
\begin{equation*}
\sum_{k\in \Sigma_n}t_k(j,n)u(x-k)=w(x-j)\geq \chi_j(x)
\end{equation*}
for all $x\in\L_n$. Moreover, the $\ell^1$-norms of solutions of
\eqref{e-Loesung} are uniformly bounded in $j$ and $n$. Therefore, the
vectors $t(j,n)$ satisfy Hypothesis {\ref{hyp4}}.
\end{proof}

Applying Proposition \ref{Proposition:3.1} we obtain the following

\begin{corollary}
\label{cor:4.2} Assume there is a polytope $\L$  satisfying Hypothesis
{\ref{hyp5}} such that for every $\tv\in\cV(\L)$ the operator $\Tv$ is
continuously invertible in $\ell^1(K_{\tv} \cap \Z^d)$. Then for all
sufficiently large $n\in\N$ the Wegner estimate
\begin{equation*}
\E\bigl\{\tr\ \EE_{H_\omega^{\L_n}}(\J) \bigr\} \leq C\, \e^{E_2}
\Var(f)\, \#\tL_n \, |\J|
\end{equation*}
holds.
\end{corollary}

In general it is not easy to decide whether the operators $\Tv$ are
continuously invertible. It seems that no general necessary and sufficient
conditions are known. However, in low dimensions there are (partial) simple
criteria (see \cite{Baxter}, \cite{Boettcher:Silbermann},
\cite{Gohberg:Ceban:67}, \cite{GohbergF-1974}).
We state without proof the following result.

\begin{proposition}\label{Prop:4.3}
Let $d=1$: For the operator $\Tv$ to be continuously
invertible in $\ell^1(K_{\tv} \cap \Z)$ it is necessary and sufficient that
$s_a(\theta)\neq 0$ for all $\theta\in\T$ and $\wn(s_a)=0$.

Let $d=2$: For the operator $\Tv$ to be continuously invertible in
$\ell^1(K_{\tv} \cap \Z^2)$ it is necessary that $s_a(\theta)\neq 0$ for
all $\theta\in\T^2$ and $\wn(s_a)=0$.
\end{proposition}

That in the case $d=2$ the conditions $s_a(\theta)\neq 0$ for all
$\theta\in\T^2$ and $\wn(s_a)=0$ are not sufficient for the invertibility
of $\Tv$ follows from the following well-known example (see
\cite{Douglas:Howe}):
\begin{equation*}
s_a(\theta_1,\theta_2) = 16\, \e^{2i\theta_1}\, \e^{-2i\theta_2} - 36\,
\e^{i\theta_1}\, \e^{-i\theta_2} + 27\, \e^{-i\theta_1}\, \e^{i\theta_2},
\qquad (\theta_1,\theta_2)\in\T^2,
\end{equation*}
where the cone $K_{\tv}$ is chosen to be  the quarter-plane $(\R_+)^2$. A
number of sufficient conditions can be found in the book
\cite{Boettcher:Silbermann}.

The following result is a criterion for the Lipschitz continuity of the IDS
if $d \leq 2$.

\begin{theorem}[implies Theorem \ref{DOStheorem}]
\label{leq:2} Let $d\leq 2$. Assume that $s_a(\theta)\neq 0$ for all
$\theta\in\T^d$. Then for all sufficiently large $l$ there exists a
constant $C$ independent of $l$ and $\J$ such that
\begin{equation*}
\E \{ \tr\ \EE_\omega^l(\J) \} \leq C \; \e^{E_2} \Var(f)  |\J| \ l^d.
\end{equation*}
\end{theorem}

Recall that $\Lambda_l(0)=(-l/2,l/2)^d$ denotes a cube centered at the
origin, $H_\omega^{\Lambda_l(0)}$ the restriction of the operator
$H_\omega$ to the cube $\Lambda_l(0)$ with Dirichlet or periodic boundary
conditions, and $\EE_\omega^l:=\EE_{H_\omega^{\Lambda_l(0)}}$ the spectral
family of the operator $H_\omega^{\Lambda_l(0)}$.

Theorem \ref{leq:2} implies that the IDS is Lipschitz continuous. Thus, the
density of states $n(E)=\mathrm{d} N(E)/\mathrm{d} E$ exists for
a.e.~$E\in\R$ and is locally uniformly bounded.

\renewcommand{\proofname}{Proof of Theorem \ref{leq:2}}

\begin{proof}
Translating the convolution vector $a$ if necessary we may assume by \eqref{umdreh} that $\wn(s_a)=0$.

The case $d=1$ is proven by combining the results of Corollary
\ref{cor:4.2} and Proposition \ref{Prop:4.3}.

Assume that $d=2$. By the Kozak-Simonenko theorem \cite{Kozak:Simonenko:80}
(see Theorem \ref{Theorem:B.3} in Appendix \ref{app:A} below) there is a
family $\Pi_n\subset\R^2$ of finite convex polygons whose vertices belong to
$\Z^2$ and whose angles are all close to $\pi$ such that $\Tn$ with
$\Sigma_n=\Pi_n\cap\Z^d$ is continuously invertible in $\ell^1(\Sigma_n)$
and $\|\Tn^{-1}\|_{1,1}$ is bounded uniformly in $n\in\N$. Thus, for every
$n$ and all $j\in\Sigma_n$ the equation
\begin{equation*}
\Tn \ t(j,n) = \delta_j
\end{equation*}
has a solution and its $\ell^1$-norm is bounded uniformly in $n$, i.e.~the
vector $t(j,n)$ satisfies the conditions of Hypothesis {\ref{hyp4}}.
Moreover, the polygons $\Pi_n$ are monotone increasing and tend to $\R^d$.

For any given $l\in\N$ choose an $n\in\N$ such that
$\widetilde{\Lambda}_l(0) \subset \Sigma_n$.
Noting that both conditions of Hypothesis {\ref{hyp4}} are
satisfied for cubes $\Lambda_l(0)$, Proposition \ref{Proposition:3.1}
implies
\begin{equation*}
\E\bigl\{\tr\ \EE_\omega^l(\J) \bigr\} \leq C\ \e^{E_2} \Var(f)\ |\J| \
|\Lambda_l(0)|
\end{equation*}
with some $C>0$.
\end{proof}

\begin{remark}
{\rm In the case $d=2$ the conditions $s_a(\theta)\neq 0$ for all
$\theta\in\T^2$ and $\wn(s_a)=0$ is sufficient for the invertibility of the
half-plane Toeplitz operators on $\ell^1(\Z\times\Z_+)$. Kozak and Simonenko
implicitly use this fact in \cite{Kozak:Simonenko:80} to construct polygons
$\Pi_n$ such that any $\Tv$ is almost a half-plane Toeplitz operator.}
\end{remark}

\smallskip

We turn now to

\renewcommand{\proofname}{Proof of Theorem \ref{WEtheorem}}

\begin{proof}
Since $s_a$ is sectorial we may assume without loss of generality that
$\Re\, s_a (\theta) \geq 0$ for all $\theta\in\T^d$. This condition implies
that $T_l$ is invertible as a map from $\ell^2$ to itself for all $l\in\N$
\cite{Boettcher:Grudsky:multi}, \cite{Boettcher:Grudsky}.

First, consider the case when $\Re\, s_a$ has exactly $M\geq 1$ pairwise
different (not necessarily simple) zeros on $\T^d$, which we denote by $z_m
=(\theta_1^m, \dots, \theta_d^m)$, $m=1,\dots,M$. Let $\delta>0$  be such
that the balls $B_\delta(z_m)\subset\T^d$ are disjoint and set
\begin{equation}\label{def:Da}
\begin{split}
D_m(n)^{-1} & = \inf\{\Re\, s_a(\theta) | \, n^{-1} \leq \|\theta-z_m\|_2
\leq \delta\},\\ \tD(n) & =\max_{m=1}^M D_m(n).
\end{split}
\end{equation}
By estimates obtained by B\"{o}ttcher and Grudsky in Section 8 of
\cite{Boettcher:Grudsky:multi} (cf.\ Theorem 3.4 in
\cite{Boettcher:Grudsky} for the case $d=1$) we have
\begin{equation}\label{vor:4.3}
\|T_l^{-1}\|_{2,2}\leq\; C\; \tD(2\cdot 13^{dM} l)
\end{equation}
with a constant $C>0$ depending on the symbol $s_a$ only. Here
$\|\cdot\|_{2,2}$ denotes the norm of a linear map from $\ell^2$ to itself.
This implies for the $\ell^1\to \ell^1$-norm
\begin{equation}\label{kom:1}
\|T_l^{-1}\|_{1,1} \leq\; C \; D_a(l) \quad \text{ with }\quad D_a(l):=
l^{d/2} \tD(2\cdot 13^{d M} l).
\end{equation}
Using the fact that $\Re\;s_a$ is a trigonometric polynomial with a finite
number of zeros, from \eqref{def:Da} for sufficiently large $n$ we obtain
\begin{equation*}
\tD(n) \leq C n^{\rho}
\end{equation*}
with $\rho\in\N$ the maximal order of the zeros $z_m$. Combining this with
\eqref{kom:1} proves the claim for the case when $\Re\, s_a$ has $M\geq 1$
zeros.

Now we turn to the case $\Re\, s_a(\theta)>0$ for all $\theta\in\T^d$.
Again since $\Re\, s_a$ is a trigonometric polynomial, there is a number
$\mu>0$ such that $\Re\, s_a(\theta)>\mu$ for all $\theta\in\T^d$.
Therefore (see Section 8 in \cite{Boettcher:Grudsky:multi}), the estimate
\eqref{vor:4.3} holds with $\widetilde{D}_a\equiv 1$.
\end{proof}

\section{Localization For Potentials With Small Negative Part\label{LocSec}}

In this section we give a proof of Theorem \ref{locthm}. A box
$\Lambda_l(0)$ is called \emph{$E$-suitable for} $H_\omega$ if $l \in 6
\N$, $E \notin \spec(H_\omega^l)$, and
\begin{equation*}
\| \chi^{out} (H_\omega^l-E)^{-1} \chi^{in} \| \leq l^{-2bd}.
\end{equation*}
Here $\chi^{out}$ denotes the characteristic function of the boundary belt
$\Lambda_{l-1}(0)\setminus\Lambda_{l-3}(0)$ and $\chi^{in}$ the
characteristic function of the interior box $\Lambda_{l/3}(0)$.

Applying Corollary 3.12 in \cite{Germinet:Klein:01a} we obtain the
following result:

\begin{theorem}
\label{GKthm} There exists a number $l_1\in\N$ such that if for some
$\tilde{l}\geq l_1$ we can verify the inequality
\begin{equation}
\label{GKini} \PP\{\Lambda_{\tilde l}(0)\; \text{is not $E$-suitable for
$H_\omega$} \} \leq 841^{-d}
\end{equation}
for all $E$ in some compact interval $I\subset\R$ and the inequality
\begin{equation*}
\E \bigl\{ \tr\ \EE_{H_\omega^l}([E -\epsilon,E +\epsilon])\bigr\} \leq C\,
\epsilon\; l^{bd} \quad \text{ for all  $\epsilon \in [0,1]$  and all $l\ge
\tilde{l}$},
\end{equation*}
for all $E$ in some open interval containing $I$,
then for any compact $K \subset \R^d$ and any $q>0$
\begin{equation*}
\E \left\{\sup_{\|\varphi \|_\infty \leq 1}  \left\| |X|^{q/2}
\varphi(H_\omega) \EE_{H_\omega}(I) \chi_K \right\|_{HS}^2 \right\} <
\infty,
\end{equation*}
i.e., strong Hilbert-Schmidt dynamical localization in the energy interval
$I$ holds for $H_\omega$.
\end{theorem}

\begin{remark}\label{r-GK}
The scale $l_1$ in Theorem \ref{GKthm} depends on the single site potential
only through its support and $L^p$-norm (see Theorem 3.4 in
\cite{Germinet:Klein:01a} and Theorem A.1 in \cite{Germinet:Klein:01b}).

Let $U \subset \R^d$ be a fixed bounded set. In the sequel we consider only
such single site potentials $u$ whose non-positive part $u_-$ is supported
in $U$. Under this assumption, for any generalized step function $u$ and
any finite  constant $\tilde{\gamma}>0$ the $L^p$-norm of $u$ can be
bounded independently of  $u_-$ provided that
$\|u_-\|_\infty<\tilde{\gamma}$.
\end{remark}

Let $u= u_+ +u_-$ with $u_+\ge 0$ and $u_-\le 0$ be the decomposition of the
given single site potential $u$ in its non-negative and non-positive parts
such that
\begin{equation*}
H_\omega = H_{\omega,+} + \sum_{k \in \Z^d} \omega_k u_-(\cdot - k),
\end{equation*}
where $H_{\omega,+}$ is the family of Schr\"{o}dinger operators \eqref{Hplus}
with the non-negative single site potential $u_+$. Below we will show that
condition \eqref{GKini} in Theorem \ref{GKthm} is fulfilled for the
operator $H_\omega$ provided that $\|u_-\|_\infty$ is sufficiently small and $\supp u_-\subset U$.

\smallskip

By Hypothesis {\ref{ISEass}} $E_0$ is a lower spectral edge of the spectrum
of the operator $H_{\omega,+}$ for $\PP$-almost all $\omega\in\Omega$.
In other words $E_0$ is in the spectrum of $H_{\omega,+}$ for almost every $\omega$
and there is an $a<E_0$ such that the interval $(a,E_0)$ is in the resolvent
set of $H_{\omega,+}$.

\begin{lemma}
Let $B=B_+ - B_-$ be a  bounded, selfadjoint operator with $B_\pm\geq 0$.
Assume that an interval $(a,b)$ belongs to the resolvent set of a
self-adjoint operator $A$. If $\|B_+\|+\|B_-\|<(b-a)$, then the interval
$(a+\|B_+\|,b-\|B_-\|)$ is in the resolvent set of the operator $A+B$.
\end{lemma}

\renewcommand{\proofname}{Proof}

\begin{proof}
Let $P$ and $Q$ be the spectral projections for the operator $A$ associated
with $(-\infty, a]$ and $[b,\infty)$, respectively, such that $A=PAP +
QAQ$. Obviously, $PAP + PBP \leq (a + \|B_+\|)P$ and $QAQ + QBQ \geq (b -
\|B_-\|)Q$. Therefore, the interval $(a+\|B_+\|,b-\|B_-\|)$ belongs the
resolvent set of the operator $A + PBP + QBQ$. Observing that the
perturbation of this operator by $PBQ+QBP$ does not diminish the length of
the spectral gap (see Theorem 2.1 in \cite{AdamyanL-95}), we obtain the
claim.
\end{proof}

We will use the lemma several times in the sequel without explicit
reference.

Since $\supp u_-\subset U$, there exists a constant $c_g>0$ such that
\begin{equation*}
0 \ge \sum_{k \in \Z^d} \omega_k u_-(x-k) \ge -c_g \omega_+ \|u_-\|_\infty.
\end{equation*}
Thus, the potential $\sum_{k\in\Z^d} \omega_k u_-(\cdot-k)$ is a bounded
operator with norm bounded by\newline $c_g\,\omega_+\, \|u_-\|_\infty$.

Let $2 C_{\mathrm{gap}}$ be the length of the spectral gap of $H_{\omega,+}$
below $E_0$,
\begin{equation*}
2 C_{\mathrm{gap}} :=\dist\left(E_0,
\sup(-\infty,E_0)\cap\spec(H_{\omega,+})\right).
\end{equation*}

Choose $\gamma_1\leq C_{\mathrm{gap}}/c_g \omega_+$. Thus, if
$\|u_-\|_\infty \leq \gamma_1$, then the interval $[E_0 - C_{\mathrm{gap}},
E_0]$ contains a lower spectral edge $E_1$ of the operator $H_\omega$.
Since the potential $\sum_{k \in \Z^d} \omega_k u_-(\cdot - k)$ is
non-positive, the spectral gap of the operator $H_\omega$ below $E_1$ has
length at least $C_{\mathrm{gap}}$.

Consider the $(l\mathbb{Z})^d$-periodic approximations  of $H_{\omega}$ and
$H_{\omega,+}$
\begin{equation}
\label{e-PerApprox} H_{\omega,l} = H_0 + \sum_{k \in \Z^d}
\omega_{\displaystyle \hat{k}}u(\cdot - k), \quad H_{\omega,+,l} = H_0 +
\sum_{k \in \Z^d} \omega_{\displaystyle \hat{k}}u_+(\cdot - k)
\end{equation}
with
\begin{equation*}
\hat{k}\in (\Z_l)^d,\qquad \hat{k} = k \mod (l\mathbb{Z})^d.
\end{equation*}

The following lemma infers the initial scale estimate \eqref{GKini} from an
estimate on the distance between the spectrum of the operator $H_\omega^l$
and the reference energy $E_0$.

\begin{lemma}\label{inilem}
Fix $\xi>0$ and $\beta\in (0,1)$. Then there exist an $l_2 \in \mathbb{N}$
and a positive number $\gamma_2\leq\gamma_1$ such that if the inequality
\begin{equation}
\label{sparse} \PP \left\{ \dist\bigl(\spec(H_\omega^l),E_0\bigr) \leq
l^{2(\beta-1)}/2  \right\} \leq l^{-\xi}
\end{equation}
holds for some $l \geq l_2$, then the estimate \eqref{GKini} with
$\tilde{l}=l$  holds for all
\begin{equation*}
E\in I := [E_1 ,\; E_0 +l^{2(\beta-1)}/4].
\end{equation*}
whenever $\|u_-\|_\infty \leq \gamma_2$.
\end{lemma}

\begin{proof}
Choose $l$ sufficiently large such that $l^{2(\beta-1)}/2 \le
C_{\text{gap}}$. If we restrict the operator $H_{\omega,l}$ to a cube
$\Lambda_l$ using periodic boundary conditions, the spectrum of the
resulting restriction $H_\omega^l$ is contained in $\spec(H_\omega)$
almost surely, see (1.1) in \cite{Kirsch:Stollmann:Stolz} and Remark 5.2.2
in \cite{Veselic-2002a}. Hence, the length of the spectral gap below $E_1$
is not diminished by imposing periodic boundary conditions. Choose $\gamma_2
\leq l^{2(\beta-1)}/2 c_g \omega_+$ and assume that $\|u_-\|_\infty\leq
\gamma_2$. Then $E_1 \geq E_0 -l^{2(\beta-1)}/2$. Thus, for a subset of
$\Omega$ of measure at least $1-l^{-\xi}$, the interval $[E_-, E_0+
l^{2(\beta-1)}/2]$ with
$E_-:=\sup\left((-\infty,E_1)\cap\spec(H_\omega)\right)$ contains no
spectrum of $H_\omega^l$.

We use the Combes-Thomas estimate \cite{Barbaroux:Combes:Hislop},
\cite{Combes:Thomas} to deduce the decay estimate of the sandwiched
resolvent in \eqref{GKini} from the assumption \eqref{sparse} on the
distance between $E\in I$ and the spectrum. We use a formulation of this
bound as it is given in Theorem 2.4.1 in \cite{Stollmann-2001}. It suits
our purposes because there the dependence of the constants on the
quantities we are interested in is explicitly given.  It implies
\begin{equation}
\label{CTe} \|\chi^{out} (H_\omega^l-E)^{-1} \chi^{in} \| \leq C \,
l^{2-2\beta} \exp \left (-\frac{1}{C} \sqrt{C_{\text{gap}}} \, l^{\beta}
\right)
\end{equation}
for all $E \in [E_1, E_0+ l^{2(\beta-1)}/4]$, some $C>0$, and all $\omega$
in a subset of measure greater or equal to $1-l^{-\xi}$. The dependence of
the constant $C$ on $V_\omega$ is through $\sup_x \|\chi_{\Lambda_1(x)}
V_\omega\|_{L^p}$ only. Thus, this  constant can
be chosen uniformly in $u_-$, whenever $\|u_-\|_\infty$ is uniformly bounded
and $\supp u_- \subset U$ as assumed before.

Choose $l_2$ sufficiently large so that the r.h.s.~of \eqref{CTe} is
bounded by $l^{-2bd}$ for all  $l\ge l_2$. The scale $l_2$ depends on $d$,
$C$, $C_{\text{gap}}$, $\beta$, and $b$ only. Thus, it can be chosen
independently  of $\|u_-\|_\infty$ as long as $\|u_-\|_\infty<\gamma_2$ and
$\supp u_- \subset U$. If necessary, enlarge $l_2$ such that $l_2 \ge
841^{d/\xi}$, and, thus, the probability estimate in \eqref{GKini} becomes
valid.
\end{proof}

\begin{lemma}
\label{PDsparse} Under Hypothesis {\ref{ISEass}} there exist $\xi>0$, $\beta
\in (0,1)$, and $l_3 \in\N$  such that for all $l \geq l_3$ the inequality
\begin{equation}
\label{PosSparse} \PP \bigl\{ \dist\bigl(\spec(H_{\omega,+}^l), E_0\bigr)
\leq l^{2(\beta-1)} \bigr\} \le l^{-\xi}
\end{equation}
holds.
\end{lemma}

\begin{proof}
Note that, since the random potential of $H_{\omega,+}$ is non-negative,
and $\min \supp f=0$, $E_0$ is a common lower spectral edge of both the
unperturbed, periodic operator $H_0$ and of the random one $H_{\omega,+}$.
Theorem 2.2.1 in \cite{Stollmann-2001} proves \eqref{PosSparse} under
assumption (ii) and Proposition 1.2 in \cite{Veselic-2002b} under
assumption (iii). In the energy/disorder regime (i) and the additional
assumption that $\supp f$ is an interval, the estimate is proven in
Proposition 4.2 of \cite{Kirsch:Stollmann:Stolz}. If the support of $f$ has
several components we still have Lifshitz tails at the bottom of the
spectrum and the statement of Proposition 1.2 in \cite{Veselic-2002b} holds
with the same proof.
\end{proof}

\begin{lemma}\label{NegPert}
If $\|u_-\|_\infty \leq l^{2(\beta-1)}/ 2c_g \omega_+$ for some $l\in\N$
and $\beta\in(0,1)$, then
\begin{equation*}
\PP \{  \dist(\spec(H_\omega^{l}),  E) \leq l^{2(\beta-1)}/2 \} \leq \PP
\{ \dist(\spec(H_{\omega,+}^{l}) ,E) \leq l^{2(\beta-1)} \}
\end{equation*}
for any $E \in \R$.
\end{lemma}

\begin{proof}
By the min-max Theorem for eigenvalues
\begin{equation*}
H_{\omega,+}^{l} - l^{2(\beta-1)}/2 \leq H_{\omega,+}^{l} - c_g
\omega_+
 \leq H_\omega^{l} \leq H_{\omega,+}^{l}
\end{equation*}
in the sense of quadratic forms. This implies the inclusion
\begin{equation*}
\{ \omega | \, \dist(\spec(H_\omega^{l}),  E) \leq l^{2(\beta-1)}/2 \}
\subset \{ \omega | \, \dist(\spec(H_{\omega,+}^{l}) ,E) \leq
l^{2(\beta-1)} \}.
\end{equation*}
\end{proof}

\renewcommand{\proofname}{Proof of Theorem \ref{locthm}}

\begin{proof}
Fix $U \subset \R^d$ as in Remark \ref{r-GK},
$\xi>0$, $\beta\in(0,1)$, and $l_3\in\N$ as in Lemma \ref{PDsparse},
and $l_2\in\N$ and $\gamma_2>0$ as in Lemma \ref{inilem}. Assuming
$\|u_-\|_\infty\leq \tilde{\gamma}$ choose $l_1\in\N$ as in Theorem
\ref{GKthm} uniformly in $\|u_-\|_\infty$ (see Remark \ref{r-GK}).

Let $l_4$ be the smallest number in $6 \N$ such that $l_4 \geq \max
\{l_1,l_2,l_3\}$. Set
\begin{equation*}
\gamma=\min\{l_4^{2(\beta-1)}/2c_g\omega_+,\tilde{\gamma}, \gamma_2\}.
\end{equation*}
Then, if $\|u_-\|_\infty\leq \gamma$, Lemmata \ref{inilem}, \ref{PDsparse},
and \ref{NegPert} imply that the inequality \eqref{GKini} holds for
$\tilde{l}=l_4$. Now, Theorem \ref{GKthm} implies Theorem \ref{locthm}.
\end{proof}

\section{H\"{o}lder Continuity of the IDS}\label{s-HK}

In this section we revisit the main result of the paper \cite{Hislop:Klopp}
by Hislop and Klopp - the H\"{o}lder continuity of the IDS below the spectrum
of the operator $H_0$ for sign-indefinite single site potentials. Assuming
that the density of the conditional probability distribution is piecewise
absolutely continuous (see Hypothesis (H4) in \cite{Hislop:Klopp}) Hislop
and Klopp proved the finite-volume H\"{o}lder-Wegner estimate (Theorem 1.1)
with linear dependence on the volume of the domain. This estimate
immediately implies that the IDS is H\"{o}lder continuous.

We will prove that this hypothesis on the density of the conditional probability
distribution can be relaxed: It suffices to require that this density is of
bounded total variation.

We will need the following elementary

\begin{lemma}
\label{l-ext} Let $\phi$ and $g$ be real-valued functions of bounded variation
on the interval $[m,M]$. Assume that $\phi(m)=0$ and $g$ is continuous. Then
\begin{eqnarray*}
\left\vert\int_{m}^{M} \phi(t) \mathrm{d}g(t)\right\vert \leq 2\Var (\phi) \,
\|g\|_\infty
\end{eqnarray*}
with $\Var(\phi)$ the variation of $\phi$ on the interval $[m,M]$.
\end{lemma}

\renewcommand{\proofname}{Proof}

\begin{proof}
Using the integration by parts formula for Riemann-Stieltjes integrals we
obtain
\begin{equation*}
\begin{split}
\left \vert \int_m^M \phi(t) \mathrm{d}g(t) \right \vert &=\; \left \vert \phi(M)
g(M) - \int_m^M  g(t) \mathrm{d}\phi(t) \right \vert
\\
&\leq \; \left \vert \phi(M) g(M)\right \vert  + \Var (\phi) \, \| g \|_\infty \\
&\leq \; \|\phi\|_\infty\; \|g\|_\infty  + \Var (\phi) \; \| g \|_\infty.
\end{split}
\end{equation*}
Observing that $\|\phi\|_\infty\leq \Var(\phi)$ completes the proof.
\end{proof}

A similar idea is applied to H\"{o}lder continuous measures in
\cite{HundertmarkKNSV}.

\begin{hypothesis}
\label{h-HK}
Let $H_\omega = H_0 + V_\omega$, where $H_0= -\Delta +V_0$ and $V_0$ is a
$\Z^d$-periodic potential infinitesimally bounded with respect to $\Delta$.
The single site potentials $\{u_k\}_{k\in\Z^d} \subset C_0(\R^d)$ of
\begin{equation*}
V_\omega (x)= \sum_{k \in\Z^d} \omega_k \, u_k(x-k)
\end{equation*}
do not vanish at the origin $0 \in \R^d$ and satisfy uniformly the bound
\begin{equation*}
\sum_{k \in\Z^d} \|u_k \|_{L^p (\Lambda_1(k))} \le C_u < \infty.
\end{equation*}
The conditional probability measures $\mu_k$ of $\omega_k$
with respect to $\omega^{\perp k}:= \{\omega_j|\, j\in \Z^d, j \neq k\}$
admit a conditional density $f_k(\omega^{\perp k},\cdot) \in L_c^\infty
(\R) $ such that
\begin{equation}\label{unifVar} \sup_{k \in\Z^d} \sup_{\omega^{\perp k}} \Var \,
[f_k(\omega^{\perp k},\cdot) ] \le  C_f <\infty
\end{equation}
and there exist $ m,M \in \R$ such that $\supp
f_k(\omega^{\perp k},\cdot) \subset [m,M]$ for all values of $\omega^{\perp
k}$ and all $k \in\Z^d$.
\end{hypothesis}

If the $\{\omega_k\}_{k \in\Z^d}$ form an i.i.d.~sequence, condition
\eqref{unifVar} simplifies to $ \Var (f_0)   < \infty$. Note that the
condition on the bounded total variation of the densities $f_k$ is in
particular satisfied if they are piecewise absolutely continuous.

Here is the extension of Theorem 1.1 of \cite{Hislop:Klopp} to densities
with bounded total variation:

\begin{theorem}
\label{HK} Let $H_\omega$ satisfy Hypothesis {\ref{h-HK}} and $E_0$ be such
that $\delta:=\inf \spec(H_0) -E_0>0$. For any $q<1$ there exists $C_q \in
(0,\infty)$ such that for all $\epsilon > 0$ one has
\begin{eqnarray*}
\PP\{\omega| \, \spec(H_\omega^l) \, \cap \, [E_0 -\epsilon, E_0 ] \neq
\emptyset \} \le C_q \,\epsilon^q |\Lambda_l|.
\end{eqnarray*}
The constant $C_q$ depends only on $d$, $q$, the periodic background
potential $V_0$, the conditional measures $\mu_k$, the single site potentials
$u_k$, and the distance $\delta$.
\end{theorem}

\begin{proof}
We follow the arguments of the proof of Theorem 1.1 in \cite{Hislop:Klopp}
up to the bound (3.15) there, which applies only to compactly supported,
bounded, locally absolutely continuous densities. Like in
\cite{Hislop:Klopp}, we assume for simplicity of notation that we are in
the i.i.d.~case and $m=0$.

We estimate the l.h.s.~of (3.15) using Lemma \ref{l-ext} with
$\phi(\lambda):=\lambda f(\lambda)$ compactly supported and
\begin{equation}
\label{ana} g\colon \omega_k \mapsto \sum_n \rho(E_n^\Lambda (\omega)),
\end{equation}
where $E_n^\Lambda (\omega)$ denote the eigenvalues of the compact
Birman-Schwinger type operator $\Gamma(\omega_k)=\sum_{j\in\tL} \omega_j \,
(H_0-E)^{-1/2} \, u_j \, (H_0-E)^{-1/2}$. Here $E$ is an energy below the
spectrum of $H_0$ and $\rho\in C_0^\infty(\R\setminus\{0\})$. The operator
$\Gamma$ depends analytically on $\omega_k$ and so do its eigenvalues.
Since $\rho$ is a smooth function and there are only finitely many
eigenvalues in the support of $\rho$, the sum in \eqref{ana} consists of
finitely many terms and the assumptions of Lemma \ref{l-ext} on $g$ are
satisfied.

The rest of the proof goes through as in Section 3 of \cite{Hislop:Klopp}
with the constant\footnote{Note that there are misprints in (3.15) and
(3.16) of \cite{Hislop:Klopp}: The $L^\infty$-norm
$\|\tilde{h}_0^\prime\|_\infty$ has to be replaced there by the $L^1$-norm
$\|\tilde{h}_0^\prime\|_1$.} $\max\{\|\phi'\|_1,$ $\phi(M)\}$ replaced by
$\Var(\phi)$.
\end{proof}

\begin{appendix}

\section{Spectral Averaging Theorem}\label{appendix}

Here we give an alternative proof of Theorem \ref{Combes-Hislop}. This
proof exhibits a relation of the spectral averaging to the theory of the
spectral shift function and, in particular, to the operator-valued version
of the Birman-Solomyak formula \cite{Gesztesy:Makarov:Naboko}.

First we need the following

\begin{lemma}\label{inequality}
Let $A_1,A_2$, and $C$ be bounded operators. If $A_1$ and $A_2$ are
self-adjoint, non-negative and satisfy $A_2^2\geq A_1^2\geq 0$ then
\begin{equation*}
\| A_1CA_1\| \leq \|A_2CA_2\|.
\end{equation*}
\end{lemma}

\begin{proof}
For $t>0$ consider
\begin{equation*}
\begin{split}
& \|(A_1+t) C (A_1+t)\|\\ & \qquad \leq\; \|(A_1+t) (A_2+t)^{-1}\|\cdot
\|(A_2+t) C (A_2+t)\|\cdot\|(A_2+t)^{-1} (A_1+t)\|.
\end{split}
\end{equation*}
Note that $(A_2+t)\geq t>0$ has a bounded inverse and
\begin{equation}\label{stern}
\begin{split}
\|(A_1+t) (A_2+t)^{-1}\|^2 & =\;\|(A_2+t)^{-1} (A_1+t)\|^2\\
&=\; \|(A_2+t)^{-1} (A_1+t)^2(A_2+t)^{-1}\|.
\end{split}
\end{equation}
{}From the assumption $A_1^2\leq A_2^2$ by the Heinz-L\"{o}wner inequality
\cite{Heinz}, \cite{Loewner} it follows that $A_1\leq A_2$ and, therefore,
$(A_1+t)^2\leq (A_2+t)^2$. Thus, the r.h.s.~of \eqref{stern} is bounded by
one. This proves the inequality
\begin{equation*}
\|(A_1+t) C (A_1+t)\| \leq \|(A_2+t) C (A_2+t)\|
\end{equation*}
for any $t>0$. Both norms are continuous in $t$. Taking the limit
$t\downarrow 0$ completes the proof of the lemma.
\end{proof}

\renewcommand{\proofname}{Proof of Theorem \ref{Combes-Hislop}}

\begin{proof}
First we pull the density $g$ out of the integral
\begin{eqnarray*}
\lefteqn{\left\|\int_{t_1}^{t_2} g(s) B \EE_{H(s)}(J) B \mathrm{d}s\right\|
= \sup_{\|\phi\|=1} \int_{t_1}^{t_2} g(s) \langle\phi, B \EE_{H(s)}(J)
B\phi\rangle \mathrm{d}s}\\ &\leq & \|g\|_\infty \sup_{\|\phi\|=1}
\int_{t_1}^{t_2} \langle\phi, B \EE_{H(s)}(J) B\phi\rangle \mathrm{d}s =
\|g\|_\infty \left\|\int_{t_1}^{t_2} B \EE_{H(s)}(J) B \mathrm{d}s\right\|.
\end{eqnarray*}
Now we write
\begin{equation*}
\int_{t_1}^{t_2} B\EE_{H(s)}(J) B\; \mathrm{d}s = B\int_{t_1}^{t_2}
\EE_{H(s)}(J) \mathrm{d}s\;B
\end{equation*}
and apply Lemma \ref{inequality} with $A_1=B$, $A_2=\kappa^{-1/2} V^{1/2}$,
and $C=\int_{t_1}^{t_2} \EE_{H(s)}(J) \mathrm{d}s$, thus, obtaining
\begin{equation}\label{step:1}
\left\|\int_{t_1}^{t_2} B \EE_{H(s)}(J) B \mathrm{d}s\right\| \leq
\kappa^{-1} \left\|\int_{t_1}^{t_2} V^{1/2} \EE_{H(s)}(J) V^{1/2}
\mathrm{d}s\right\|
\end{equation}

The rest of the proof follows the line of \cite{Gesztesy:Makarov:Naboko}.
For an arbitrary invertible dissipative operator $T$ we define its
logarithm via
\begin{equation*}
\log(T)=-i\int_0^\infty \mathrm{d}\lambda
\left((T+i\lambda)^{-1}-(I+i\lambda)^{-1} \right)
\end{equation*}
with $I$ the identity operator.

We claim that the equality
\begin{equation}\label{zeile}
\int_{t_1}^{t_2} V^{1/2}(H(s)-z)^{-1}V^{1/2} \mathrm{d}s = \log(I+(t_2-t_1)
V^{1/2}(H(t_1)-z)^{-1}V^{1/2})
\end{equation}
holds for all $z\in\C$ with $\Im\, z>0$. For $r>0$ we set
$\C_{r,+}=\{z\in\C|\ \Im\, z>r\}$. For sufficiently large $r$ and all
$z\in\C_{r,+}$
\begin{equation*}
V^{1/2}(H(s)-z)^{-1}V^{1/2} = \sum_{k=0}^\infty (t_1-s)^k
V^{1/2}(H(t_1)-z)^{-1} \left[V (H(t_1)-z)^{-1} \right]^k V^{1/2}
\end{equation*}
in the operator norm. Integrating this expression with respect to $s$ we
obtain
\begin{equation*}
\begin{split}
\int_{t_1}^{t_2} V^{1/2}(H(s)-z)^{-1}V^{1/2} \mathrm{d}s  &=
\sum_{k=1}^\infty\frac{(-1)^{k+1}}{k}
(t_2-t_1)^k \left[V^{1/2}(H(t_1)-z)^{-1}V^{1/2}\right]^k\\
&= \log(I+(t_2-t_1) V^{1/2}(H(t_1)-z)^{-1}V^{1/2}).
\end{split}
\end{equation*}
Since $I+(t_2-t_1) V^{1/2}(H(t_1)-z)^{-1}V^{1/2}$ is an invertible
dissipative operator and since the l.h.s.\ of \eqref{zeile} is analytic in
$z$ for all $\Im z>0$, this proves equation \eqref{zeile} for all $\Im z>0$.

Applying now Lemma 2.8 in \cite{Gesztesy:Makarov:Naboko} to r.h.s.\ of
\eqref{zeile} we obtain that
\begin{equation}\label{zeile:2}
0\leq \Im \int_{t_1}^{t_2} \langle\phi,V^{1/2}(H(s)-z)^{-1}V^{1/2}
\phi\rangle \mathrm{d}s \leq \pi \|\phi\|^2
\end{equation}
for arbitrary $\phi$ and arbitrary $z\in\C$ with $\Im z>0$. {}From Stone's
formula it follows that
\begin{equation*}
\langle \phi, V^{1/2}\EE_{H(s)}(J)V^{1/2}\phi\rangle \leq
\frac{1}{\pi}\lim_{\epsilon\downarrow 0}\Im \int_J \langle \phi, V^{1/2}
(H(s)-\lambda-i\epsilon)^{-1}V^{1/2}\phi\rangle \mathrm{d}\lambda.
\end{equation*}
Hence, from \eqref{zeile:2} by the Fubini theorem it follows that
\begin{equation}\label{interessant}
\int_{t_1}^{t_2} \langle \phi, V^{1/2}\EE_{H(s)}(J)V^{1/2}\phi\rangle
\mathrm{d}s \leq |J| \|\phi\|^2.
\end{equation}
Combining this with \eqref{step:1} completes the proof.
\end{proof}

\begin{remark}{\rm
The integral on the l.h.s.\ of \eqref{interessant} is related to the
spectral shift operator $\Xi(\lambda)$ (see \cite{Carey},
\cite{Gesztesy:Makarov:Naboko}, \cite{Gesztesy:Makarov:99}),
\begin{equation}\label{BirSol}
\int_{t_1}^{t_2} V^{1/2}\EE_{H(s)}(J)V^{1/2} \mathrm{d}s=\int_J
\Xi(\lambda) \mathrm{d}\lambda.
\end{equation}
For trace class perturbations $V$ the trace of this operator equals
the spectral shift function for the pair of operators $(H(t_2),H(t_1))$ such
that from \eqref{BirSol} the Birman-Solomyak formula \cite{Birman:Solomyak}
follows. An application of the Birman-Solomyak formula to spectral
averaging can be found in Section 3 of \cite{Kostrykin:Schrader:2000c}.}
\end{remark}

\section{Kozak-Simonenko Polygons}\label{app:A}

A subset $M\subset\Z^2$ is called a canonical discrete half-space if $M$
and $\Z^2\setminus M$ are closed with respect to addition. A set $M$ is
called discrete half-space if there is $j\in\Z^2$ such that $M+j$ is a
canonical discrete half-space.

Let $B(x,r)\subset\R^2$ denote an open ball of radius $r$ centered at the
point $x$. By a convex lattice polygon we will understand the convex
hull in $\R^2$ of an arbitrary finite subset of $\Z^2$.

\begin{definition}\label{def:KS}
Let $\fM(r,R)$ be the set of all convex lattice polygons $\Pi$ in $\R^2$
satisfying the following conditions
\begin{itemize}
\item[(i)]{for any $x\in\R^2$ there is a discrete half-space $M$
such that  $\Pi\cap B(x,r)\cap\Z^2 = M\cap B(x,r)$,}
\item[(ii)]{$\Pi\supset B(0,R)$.}
\end{itemize}
\end{definition}

The following fact has been stated in \cite{Kozak:Simonenko:80} without
proof.

\begin{lemma}\label{lemlem}
For any $r>0$ and $R>0$ there is a sequence $\Pi_n\in\fM(r,R)$ tending to
$\R^2$.
\end{lemma}

\renewcommand{\proofname}{Proof}

\begin{proof}
First we prove that for arbitrary $r>0$ and $R>0$ the set $\fM(r,R)$ is
nonempty. Choose an arbitrary integer $q>2r$. The proof is based on the
following trivial observation: The interval $[0,1]$ contains a finite set
$S_q$  of rational numbers $\ell$ which can be represented in the form
\begin{equation}\label{ratio}
\ell=\frac{m}{n}\quad \text{with}\quad m\in\N_0,\; n\in\N\quad \text{such
that}\quad n\leq q.
\end{equation}
Ordering the elements of the set $S_q$ in increasing order we get a finite
strictly increasing sequence of rational numbers $\{\lambda_k\}_{k=1}^K$
such that $\lambda_1=0$ and $\lambda_K$=1. Set
\begin{equation*}
\tau_k = s_k\begin{pmatrix}1 \\ \lambda_k\end{pmatrix}\in \N^2
\end{equation*}
where $s_k\geq q$ is the smallest integer number such that
$s_k\lambda_k\in\N$. Take the point
\begin{equation*}
j_0=\begin{pmatrix}0 \\ 0\end{pmatrix}\in\Z^2
\end{equation*}
and consider the walk $w=\{j_0,j_1,\ldots,j_K\}$ defined by the following
recurrent relation: $j_k=j_{k-1}+\tau_k$ (see Fig.~\ref{fig:1}). Consider
the walk $\widehat{w} =
\{\widehat{j}_0,\widehat{j}_1,\ldots,\widehat{j}_K\}$ obtained from $w$ by
a translation:
\begin{equation*}
\widehat{j}_k=j_k+\begin{pmatrix} 0 \\ -j_K^{(1)} - j_K^{(2)}
\end{pmatrix},\quad\text{where}\quad j_K=\begin{pmatrix} j_K^{(1)} \\  j_K^{(2)} \end{pmatrix}.
\end{equation*}
The initial vertex $\widehat{j}_0$ of the walk $\widehat{w}$ lies on the
vertical coordinate axis. The terminal vertex $\widehat{j}_K=\begin{pmatrix} \widehat{j}^{(1)}_K \\
\widehat{j}_K^{(2)} \end{pmatrix}$ of the walk $\widehat{w}$ lies on the
diagonal such that $\widehat{j}_K^{(1)}=-\widehat{j}^{(2)}_K$.

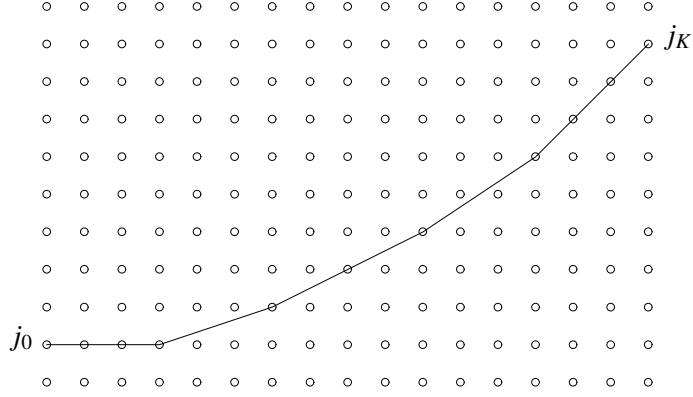
\begin{figure}[t]
\setlength{\unitlength}{1mm}
\begin{picture}(90,60)(-1,-1)
\matrixput(0,0)(5,0){17}(0,5){11}{\circle{1}}
\path(0,5)(15,5)(30,10)(50,20)(65,30)(80,45) \put(-5,5){$j_0$}
\put(82,45){$j_K$}
\end{picture}
\caption{\label{fig:1} A walk from $j_0$ to $j_{K}$ for $q=3$.\newline}
\end{figure}

Let $w^\prime=\{\widehat{j}_k\}_{k=0}^{2K}$ be the continuation of the walk
$\widehat{w}$ obtained by the mirror reflection of the walk with respect to
the diagonal,
\begin{equation*}
\widehat{j}_k= \begin{pmatrix} -\widehat{j}^{(2)}_{2K-k} \\
-\widehat{j}_{2K-k}^{(1)} \end{pmatrix},\qquad k\in\{K+1,\ldots,2K\}.
\end{equation*}
Observe that the vertices $\widehat{j}_{K-1}$, $\widehat{j}_{K}$, and
$\widehat{j}_{K+1}$ lie on the same line.

By means of mirror reflection with respect to the coordinate axes the walk
$w^\prime$ can be completed to the closed walk from $\widehat{j}_0$ to
$\widehat{j}_0$. Observe that the closed walk crosses the coordinate axes
perpendicularly. Let $\Pi\subset\R^2$ denote the convex hull of this closed
walk.

We claim that the polygon $\Pi$ satisfies condition (i). Assume first that
$x$ is a vertex of the polygon $\Pi$. Let $L_1\subset\R^2$ and
$L_2\subset\R^2$ denote the lines such that the boundary $\partial\Pi$ in a
vicinity of the vertex $x$ is a subset of $L_1\cup L_2$. Let $\cC_+$ and
$\cC_-$ denote the open cones spanned by $L_1$ and $L_2$ chosen such that
each of $\cC_+$ and $\cC_-$ touch precisely one side of the polygon $\Pi$.
By the above construction of the walk the sets
\begin{equation*}
\cC_+\cap B(x,2r)\quad\text{and}\quad\cC_-\cap B(x,2r)
\end{equation*}
do not contain points of $\Z^2$. Indeed, suppose on the contrary that either
of these sets contains a point $z\in\Z^2$. Without loss of generality we can
assume that $x\in \widehat{w}$. Then, the line passing through the points
$x$ and $z$ has a rational slope $0<m/n<1$ with $n\leq q$ such that
$m/n\neq\lambda_k$ for all $k\in\{1,\ldots,K\}$. Thus, there is a rational
number of the form \eqref{ratio} which does not belong to the set $S_{q}$. A
contradiction.

Choose an arbitrary line $L$ with a rational slope such that
$L\subset\cC_+\cup\cC_-\cup\{x\}$. The line $L$ divides $\R^2$ into two open
half-planes $\cL_L$ and $\cL_L^\prime$ such that $\cL_L$ has at least one
common point with $\Pi$ and $\cL_L\cup\cL_L^\prime\cup L=\R^2$. The set
$M_x=(L\cup\cL_L)\cap\Z^2$ is obviously a discrete half-space satisfying
condition (i).

Let now $x\in\partial\Pi$ but is not a vertex. Let $I\ni x$ be the edge of
the polygon $\Pi$, $\tv_{1,2}$ its endpoints. Set
\begin{equation*}
\begin{split}
I_0 & =\{y\in I|\; \dist(y, \tv_1) > r\quad \text{and}\quad \dist(y, \tv_2)
> r\},\\
I_1 & =\{y\in I|\; \dist(y, \tv_1) \leq r\},\\
I_2 & =\{y\in I|\; \dist(y, \tv_2) \leq r\}.
\end{split}
\end{equation*}
If $x\in I_0$, then the discrete half-space generated by the edge $I$
satisfies condition (i) of Definition \ref{def:KS}. Assume that $x\in I_1$.
The discrete half-space $M_{\tvsmall_1}$ constructed above obviously
satisfies condition (i) for the point $x$. A similar statement holds for
$x\in I_2$.

Further, assume that $x\in \Pi$ but $x\notin\partial\Pi$. If
$\dist(x,\tv)<r$ for some vertex $\tv$ of the polygon $\Pi$ we choose
$M=M_{\tvsmall}$. If $\dist(x,\partial\Pi) < r$ but $\dist(x,\tv)\geq r$
for all vertices $\tv$ of the polygon $\Pi$ we choose $M$ to be a
half-space generated by an edge $I$ having a distance to the point $x$ less
that $r$. Finally, if $\dist(x,\partial\Pi) > r$ we choose $M$ to be an
arbitrary appropriately translated half-space. In all these cases the half-space $M$ obviously
satisfies the condition (i).

Choosing $q$ sufficiently large we can satisfy condition (ii) for any given
$R>0$. Thus, the set $\fM(r,R)$ is nonempty.

Now consider an arbitrary monotone increasing sequence $\{R_n\}_{n\in\N}$
such that $R_1\geq R$ and $\displaystyle\lim_{n\rightarrow\infty}
R_n=\infty$. Let $\Pi_n\in\fM(r,R_n)$ be arbitrary. Obviously,
$\Pi_n\in\fM(r,R)$ and $\Pi_n\rightarrow \R^2$ as $n\rightarrow\infty$.
\end{proof}

Here is the main result of the paper \cite{Kozak:Simonenko:80} in the form
we need it for our application. Let $P_n$ be the projection in
$\ell^1(\Z^2)$ associated with the set $\Pi_n$,
\begin{equation*}
(P_n \phi)(k)=\begin{cases} \phi(k),& \text{if}\; k\in \Pi_n,\\
0,& \text{otherwise}. \end{cases}
\end{equation*}

\begin{theorem}\label{Theorem:B.3}
Let $T$ be a Toeplitz operator on $\ell^1(\Z^2)$ with non-vanishing symbol
$s(\theta)$. If $s(\theta)$ is a trigonometric polynomial and its
topological index is zero, then there exist positive numbers $r$, $R$, $c$
and polygons $\Pi_n\in\fM(r,R_n)$, $n\in \N$ such that the associated
projectors $P_n$ satisfy $\|(P_n T P_n)^{-1}\|\leq c$ for all $n\in\N$.
\end{theorem}

\end{appendix}

\end{document}